\documentstyle[12pt]{article}
\begin{document}
\newcommand{\NP}[1]{Nucl. \ Phys.}
\newcommand{\PL}[1]{Phys. \ Lett.}

\newcommand{\bs}{\bigskip}
\newcommand{\ba}{\begin{array}}
\newcommand{\ea}{\end{array}}
\newcommand{\eps}{\epsilon}
\newcommand{\veps}{\varepsilon}
\newcommand{\ra}{\rightarrow}
\newcommand{\la}{\leftarrow}
\newcommand{\be}{\begin{equation}}
\newcommand{\ee}{\end{equation}}

\centerline{\Large \bf   Intersecting 5-brane   solution}

\bs
\centerline{\Large \bf  of N=1, D=10 Dual Supergravity}

\centerline{\Large \bf with $\alpha'$ corrections}

\vspace*{ 1cm}

\centerline{\bf
N.A.Saulina \footnote{e-mail:  saulina@vxitep.itep.ru}
}
 \bigskip

\centerline{$  Institute \ \ of \ \ Theoretical \ \
 and \ \  Experimental \ \ Physics $ }

\bs
\centerline{ Moscow Physical Technical Institute   }

\bigskip
\begin{abstract}
 A  vacuum solution of anomaly-free
N=1, D=10 Dual Supergravity is constructed.
 This vacuum corresponds to the presence of  two 5-branes,
intersecting along $M_4$, and possesses N=1, D=4 supersymmetry.

\end{abstract}

\section{Introduction}
Various p-brane solutions in diverse supergravities were constructed
in ( \cite{Dabh} -\cite{Duff3} ),  where p-brane plays a role of a singular
source of the supergravity fields.
It was shown, that there are 2 types of
p-brane solutions of the theory with the action:
\be
                                                               \label{I0}
 I={1\over 2\,k^2} \int d^D\,x\,\sqrt{-g}\,\Bigl ( R+{1\over 2}\,
(\nabla \varphi )^2
+{e^{-\gamma \, \varphi} \over{ 2 (d+1)!}}\,F_{(d+1)}^2 \Bigr ),
\ee
where $F_{(d+1)}$ - (d+1)-form, $F_{(d+1)}=dA_{(d)}$, $\varphi$ - dilaton.
(d-1)-brane solution with  $\it electric$ charge
 $Q_E \sim \int \limits _{S^{(D-d-1)}}
e^{-\gamma \, \varphi}\,{}^\star F $ is characterized by non-zero components
of $A_{(d)}$ on the worldvolume of the brane,
while in (D-d-3)-brane solution
with $\it magnetic$ charge  $Q_M \sim \int \limits_{S^{(d+1)}}F $ only 
transverse
components of F are "alive". $S^{(N)}$ - N-sphere at infinity,
$\star$ means Hodge dual in D dimensions.

The interest in  p-brane solutions of supergravities
 is motivated by  the fact, that  in the presence of the brane-source
 ${1 \over 2}$ of initial supersymmetry is broken.
  For the review on
p-branes see ( \cite{Town}, \cite{Schw}, \cite{Stelle} ).

Using intersecting branes as a singular source of supergravity fields gives
a possibility for further breaking of supersymmetry.
In \cite{Dougl} intersecting D-branes were discussed and
it was pointed out, that in some cases they  preserve
${1\over4}$ of initial supersymmetry.

In recent works (\cite{Aref} - \cite{Roo}) a rich variety of
intersecting p-brane solutions in arbitrary dimensions were obtained
and their supersymmetry properties analyzed, putting the main attention
to solutions of D=11
Supergravity (\cite{Aref}, \cite{Pope}) and Type II supergravities
(\cite{Arg}, \cite{Tseyt}).

We want to emphasize, that so far in the literature only p-brane solutions
of  the supergravities, containing  up to $2^{d}$ order-derivative terms in
Lagrangian, were constructed.

We are interested in the solutions of
 anomaly-free ( taking into account  up to $4^{th}$ order-derivative terms )
N=1, D=10 Dual Supergravity ({\bf DS}),  which can be viewed  as
field-theory limit of a Fivebrane \cite{Strom}, \cite{D}, \cite{DL}.
The {\bf DS} lagrangian obtained in \cite{STZ1}, \cite{STZ2}
corresponds to the supersymmetrised version of $\sim \alpha' $
 anomaly cancelling Green-Schwarz (GS)  corrections \cite{GS}  to the
simple (lowest $\alpha'$-order) D=10, N=1  supergravity considered in
 \cite{CM}.
 The  problem with  the standard supergravity  is
that $\sim \alpha'$ corrections
can be made  supersymmetric only in the same $\sim \alpha' $ order.
 For complete
supersymmetrisation one must take into account the infinite number of terms
$\sim \alpha'^n, \ \ (n = 1,2,\ldots), $ containing the axionic field and
dictated by supersymmetry.
Situation is different in the dual supergravity. If the same
corrections are expressed in terms of fivebrane variables - the result
becomes exactly supersymmetric in the order $ \sim \alpha' $ , i.e. the
infinite series in $\alpha'$  is transformed to the finite number of
terms in the case of dual supergravity.

N=1, D=10  {\bf DS}   contains  6-form axionic potential $ C^{(6)}$, so that
 5-brane solution, which was  constructed for the lagrangian of \cite{STZ2}
in \cite{TZ}, is of $\it electric$ type.

In this paper we obtain a   solution of anomaly-free theory,
which  corresponds to the following situation:
 one 5-brane is in 012345, it is defined by $X^6=X^7=X^8=X^9=0$,
 the other 5-brane is in 012367 and has $X^4=X^5=X^8=X^9=0$.
The Lorentz symmetry becomes 
$SO(1,3) \times SO(2) \times SO(2) \times SO(2)$.

The solution is a vacuum configuration in the sense that all fields
depend only on coordinates $X^8$, $X^9$, which are transverse 
to both  branes, and all fermions are set to zero.

\section{Lagrangian}

The lagrangian of N=1, D=10 {\bf DS}  is equal to:
\be
                                                              \label{1}
{\cal L} =
{\cal L}^{(gauge)}+ {\cal L}^{(grav)},
\ee
where ${\cal L}^{(gauge)}$ and ${\cal L}^{(grav)} $ are lagrangians for
gauge-matter and for supergravity multiplet.

We consider   only  bosonic terms in  the lagrangian
 (\ref{1}) because we are interested in  the
 vacuum configuration. Notations correspond in general
 to \cite{STZ1}. We use the following index notations
 for flat indices:
$\hat A=(\alpha,\,A,\,\tilde A,\,a)$ , where

 $\hat A = 0,\ldots,9; $\,\, $\alpha  = 0,\ldots,3; $
  $A = 4,5; $ \,\, $\tilde A = 6,7;$\,\, $ a=8,9 $
and   for world indices:

$$X^{\hat M}=(x^{\mu},\,x^M,\,x^{\tilde M},\,y^m), $$

 $\hat M = 0,\ldots,9; $\,\,
  $\mu  = 0,\ldots,3; $
  $M = 4,5; $ \,\, $\tilde M = 6,7;$\,\,
 $ m=8,9. $ \\

We present the gravity part of the lagrangian as an expansion 
in $\alpha'$:
\be
                                                            \label{2}
{\cal L}^{(grav)}= {\cal L}^{(grav)}_0  +\alpha' {\cal L}^{(grav)}_1,
\ee
where $ {\cal L}^{(grav)}_0$ is equal to \cite{CH}
(see \cite{STZ2} for further references on the subject):
\be
                                                            \label{3}
2\,k^2\, E^{-1}\,{\cal L}^{(grav)}_0 = \phi\,\left( R-
{1\over 12}\, M_{\hat A\hat B\hat C}^2 \right).
\ee
 $ R $ is the curvature scalar, $\phi$ is the dilatonic
field, $ M_{\hat A\hat B\hat C} $ is defined by:
\be
                                                              \label{4}
 M_{\hat A\hat B\hat C} =
E_{\hat A}^{\hat P} E_{\hat B}^{\hat Q} E_{\hat C}^{\hat R}\,
 M_{\hat P\hat Q\hat R}, \ \ \
M^{(3)}={}^{\star} dC^{(6)},
\ee
where $E_{\hat M}^{\hat A}$ - tenbein.

The result for  ${\cal L}^{(grav)}_1 $ was obtained in \cite{STZ1},
\cite{STZ2}  in the form:
$$ 2\,k^2\,E^{-1}\, {\cal L}^{(grav)}_1 =  2\,R^2_{\hat A\hat B}
-R_{\hat A \hat B \hat C \hat D}^2 +
{1\over 2\cdot 6!}
\varepsilon^{\hat A \hat B \hat C\hat D\hat F_1\ldots \hat F_6}\,
{R_{\hat A \hat B}}^{\hat I \hat J}
 R_{\hat C \hat D \hat I \hat J}C_{\hat F_1\ldots \hat F_6} - $$
$$ -{1\over 2}\,R^{\hat A \hat B}(M^2)_{\hat A\hat B}
 -{1\over 6}\, M^{\hat A \hat B \hat C}D_{\hat F}^2\, 
M_{\hat A \hat B \hat C} + $$
    \be
                                                           \label{5}
+ {1\over 2}\,  M^{\hat A \hat B \hat C; \hat D}
 M^2_{\hat A \hat B \hat C \hat D}
  -{1\over 24}\, ( M^2)_{\hat A \hat B \hat C \hat D}
 M^2_{\hat A \hat C\hat B \hat D}.
\ee
 $R_{\hat A \hat B\hat C \hat D} $ is the curvature tensor,
$R_{\hat A\hat B}$ is the
Richi tensor, $(;\hat B)$ means the covariant derivative $D_{\hat B} $ 
and the following notations are introduced:

$$  M^2 =  ( M_{\hat A\hat B\hat C })^2, \ \
 ( M^2)_{\hat A\hat B} = { M_{\hat A}}^{\hat C \hat D} 
M_{\hat B \hat C \hat D} $$
$$ ( M^2)_{\hat A\hat B\hat C \hat D}=
  M_{\hat A\hat B}{}^{\hat F}  M_{\hat C \hat D \hat F}, \ \
( M^3)_{\hat A\hat B\hat C } =  M_{\hat A}{}^{\hat I \hat J}
{ M_{\hat B\hat J}}{}^{\hat K}
 M_{\hat C\hat K\hat I}. $$

We need not  ${\cal L}^{(gauge)}$ in this paper,
since we discuss a vacuum solution
with zero gauge-matter ( it is shown bellow ).

One can make fields redefinition :
\be
                                          \label{redef}
\phi =e^{{2 \over 3} \varphi}, \ \ \
g_{\hat M \hat N}=e^{-{1 \over 6} \varphi}\,g_{\hat M \hat N} {}^{(can)},
\ \ \
C^{(6)}= {C {}^{(can)} }^{(6)}
\ee
to obtain lagrangian with canonical kinetic terms
 (see for details \cite{STZ2}).
\section{Vacuum Configuration}
Let us  study equations  defining vacuum confi\-guration:
 $ <\delta_Q \, \Phi >= 0, $
 where $\Phi $ is some field but $\delta_Q$
 is a super\-sym\-metry transfor\-mation.
 When $\Phi$ is a boson, such an
equation is satisfied identically. It has nontrivial content for
 $\Phi =\psi_A,$ $\chi,$  $\lambda $, i.e. for
gravitino, dilatino and gaugino fields respectively.

 We work with formulas from \cite{STZ2} (see also \cite{BBLPT}, \cite{AFRR}
where another parametrization is used):

\be
                                                             \label{psiA}
<\delta_Q\psi_{\hat A}> = \eps_{;\hat A} +{1\over 144}
 \left( 3\, M_{\hat B \hat C \hat D}\Gamma^{\hat B \hat C\hat D}
\Gamma_{\hat A}+
\Gamma_{\hat A} M_{\hat B \hat C \hat D}
\Gamma^{\hat B \hat C \hat D}\right)\eps =0,
\ee
\be
                                                              \label{chi}
<\delta_Q \chi> = {1\over 2}\partial_{\hat A} \phi \, \Gamma^{\hat A} \eps -
\left( {\phi\over 36}\,M_{\hat A \hat B \hat C}\Gamma^{\hat A \hat B \hat C}
 -\alpha'\,
A_{\hat A \hat B \hat C}\Gamma^{\hat A\hat B \hat C}\right)\eps =0,
\ee
\be
                                                          \label{lambda}
<\delta_Q \lambda> = {1\over 4}{\cal F}_{\hat A \hat B}
\Gamma^{\hat A \hat B} \eps =0.
\ee
Here $\eps(y) $ is a 32-component Dirac spinor - the parameter of
supersymmetry transformation. It is subjected to the Majorana-Weyl
condition: $ \eps_c = {\bar\eps}\  $, $\eps = \Gamma \eps $, where
$ \Gamma $ is the chirality matrix, $\eps_c$ is
the charge-conjugated spinor. ( Let us recall that all fields depend only
on transverse coordinates $y^m$).

The 3-form field $A_{\hat A \hat B\hat C} $ in eq.(\ref{chi})
is equal to \cite{STZ1},\cite{STZ2}:

$$ A_{\hat A \hat B \hat C} = -{1\over 18} {D_{\hat F}}^2 \,
 M_{\hat A \hat B \hat C} +
{7\over 36}( M^2_{\hat D[\hat A \hat B\hat C]}){}^{;\hat D}
+{1\over 36}  M_{\hat D\hat E[\hat A;\hat B}
{ M_{\hat C]}}{}^{\hat D \hat E}- $$
$$- {5\over 8\cdot 243} M^2\, M_{\hat A\hat B\hat C}+
{5\over 8\cdot 27}  M^2_{\hat D[\hat A}{ M_{\hat B\hat C]}}{}^{\hat D}
 -{5\over 4\cdot 27} M^3_{\hat A\hat B\hat C}-$$
\be
                                                            \label{Aabc}
-{1\over 4\cdot 972}{\veps_{\hat A\hat B\hat C}}^
{\hat D\hat E \hat F \hat G\hat H\hat I\hat J} M_{\hat D\hat E\hat F}
( M_{\hat H\hat I\hat J;\hat G} + M^2_{\hat G\hat H\hat I\hat J}).
\ee

( We  also hold on the standard notation:
 $ \Gamma_{\hat A_1\ldots \hat A_k}
= \Gamma_{[\hat A_1}\Gamma_{\hat A_2} \ldots \Gamma_{\hat A_k]}. $)

We begin our study from  eq.(\ref{psiA}). It is essential thing, that to
find non-trivial solution of eq.(\ref{psiA}) one must impose two additional
conditions on $\eps$:
\be
                                                               \label{cond}
\veps_{bj}\,\Gamma^{45}\,\Gamma^j\,\eps=\nu\,\Gamma_b\,\eps, \ \ \
\nu^2=1, \ \ \
\veps_{bj}\,\Gamma^{67}\,\Gamma^j\,\eps=\tilde {\nu}\,\Gamma_b\,\eps ,
\ \ \  {\tilde {\nu}}^2=1,
\ee
thus keeping ${1\over 4} $ of initial N=1, D=10 supersymmetry.
We conclude that our vacuum  possesses N=1, D=4 supersymmetry.

Then, using  the following anzats for non-zero components of tenbein:
\be
                                                              \label{6}
{E_{\hat M} }^{\hat A} = \left(
\ba{cccc}
e^{\xi_1(y)}\,\delta_\mu^\alpha  &   & & \\
        & e^{\xi_2(y)}\, \delta_M^A& & \\
      &  & e^{\xi_3(y)}\, \delta_{\tilde M}^{\tilde A}&  \\
    & &    & e^{\xi_4(y)}\,  \delta_m^a \\
\ea
\right)
\ee
and of axionic potential:
\be
\label{axion}
C^{(6)}=\tilde{\lambda}\, e^{H_1(y)}\,dx^0\wedge\,
\ldots \wedge dx^3\wedge\,dx^4\wedge\,dx^5+
\lambda\, e^{H_2(y)}\,
dx^0\wedge\,\ldots \wedge dx^3\,\wedge \,dx^6\wedge\,dx^7,
\ee

one can obtain from eq.(\ref{psiA}):

\be
                                                             \label{xi1}
\xi_1={1 \over 6}\, (H_1 + H_2),\ \ \ \ \  \xi_2={1 \over 6}\,H_1 -
{ 1 \over 3} \,H_2,
\ee

\be
                                                             \label{xi2}
\xi_3={ 1 \over 6}\,H_2 -
 {1 \over 3} \,H_1,\ \ \ \ \
 \xi_4=-{1 \over 3} \,(H_1+H_2),
\ee
\be
\label{ll}
\lambda=\nu=\pm\,1, \ \ \ \tilde \lambda=\tilde \nu=\pm\,1,\ \ \ \
 \eps(y)= e^{-{\xi_4\over4}}\,\eps^0,
\ee
where $\eps^0 $ is a constant  spinor.

 Anzats (\ref{6}), (\ref{axion}) with account for relations
(\ref{xi1})-(\ref{ll})
leads to the following non-zero
components of $M^{(3)}$-form :

\be
                                                             \label{7}
 M_{\tilde P \tilde Q r} =\tilde{\nu} \, \veps_{\tilde P \tilde Q}\,
\delta_r^i\, \veps_{ij} \,{H_1}^j, \ \ \ \ \
 M_{PQr} =\nu \, \veps_{PQ}\,\delta_r^i \, \veps_{ij} \, {H_2}^j,
\ee

 where we introduce  the notations:
\be
\label{hh}
{H_1}^j=\eta^{jm}\,\partial_m\,H_1,\ \ \ \
{H_2}^j=\eta^{jm}\,\partial_m\,H_2, \,\,\, \eta^{jm}=-\delta^{jm}
\ee
and
\be
\label{ee}
\veps_{PQ}=e^{2\,\xi_2}\, \veps_{AB}\, \delta_P^A\, \delta_Q^B, \ \ \
\veps_{\tilde P \tilde Q}=e^{2\,\xi_3}\, \veps_{\tilde A \tilde B}\,
\delta_{\tilde P}^{\tilde A}\, \delta_{\tilde Q}^{\tilde B}.
\ee
We give formulas for surviving components of  10D-spin-connection and
curvature tensor in Appendix.

Now we turn to  eq.(\ref{lambda}), which reduces for our vacuum to
${\cal F}_{ab}\Gamma^{ ab} \eps =0$. It follows that
  $ {\cal F}_{ab}=0 $.

We are left with eq.(\ref{chi}) which is the most complicated one.
Taking into account eq.(\ref{7}) and formulas from Appendix  one is able
to calculate:
\be
                                                            \label{Ahat}
A_{\hat A \hat B \hat C}\,\Gamma^{\hat A \hat B \hat C}\,\eps =
-\left[e^{-3\xi_4}\,({{\xi_4}^f}_f -
{1\over 2}\,{\xi_4}^f\,{\xi_4}_f)\right]^{j}\,\Gamma_j\,\eps.
\ee

Using (\ref{Ahat}) one can drop one derivative
 in eq.(\ref{chi}) and find the expression for dilaton:

\be
                                                             \label{dil}
 \phi=C_0\,e^{\xi_4} +2\,\alpha'\,C_0\, e^{-2\xi_4}\,
({{\xi_4}^f}_f -
{1\over 2}\,{\xi_4}^f\,{\xi_4}_f),
\ee
where $C_0$ is an arbitrary constant.

At this stage of our consideration we expressed all fields in terms of two
arbitrary functions $H_1$ and $H_2$.

We examine bellow whether equations of motion impose
constraints on these functions.

\section{Equations of Motion}

 Equations
for N=1, D=10 {\bf DS} obtained in the lowest and next order in
$\alpha' $  are used (see \cite{STZ1}, \cite{STZ2}), see also  \cite{P}
where another parametrization was considered).

 We start from  axion EM, which can be written in the form \cite{STZ1}:

\be
                                                          \label{dHabc}
{\cal H}_{[\hat A\hat B\hat C;\hat D]}=
3\,\alpha' R_{[\hat A\hat B}{}^{\hat E\hat F}R_{\hat C\hat D]\hat F\hat E},
\ee
where

$$ {\cal H}_{\hat A\hat B\hat C} =\phi \,M_{\hat A\hat B\hat C} -2\,
\alpha'\Bigl(-{D_{\hat F}}^2\, M_{\hat A\hat B\hat C}
 +3\,(M^2_{\hat D[\hat A\hat B\hat C]})^{;\hat D} + $$
\be
                                                             \label{Habc}
+{3\over 2} M_{\hat D\hat F[\hat A;\hat B} M_{\hat C]}{}^{\hat D\hat F}
 -3\,R_{\hat D[\hat A}  M_{\hat B\hat C]}{}^{\hat D} -
{1\over 2} M^3_{[\hat A\hat B\hat C]}\Bigr).
\ee
For our configuration we have:
$R_{[\hat A\hat B}{}^{\hat E\hat F}R_{\hat C\hat D]\hat F\hat E}=0$
 and only the ${\cal H}_{ABc}$ and ${\cal H}_{\tilde A\tilde B c}$
components survive:
$$
{\cal H}_{ABc} ={{\cal H}_{ABc}}^{(0)}+\alpha'\,{{\cal H}_{ABc}}^{(1)},$$

$${{\cal H}_{ABc}}^{(0)}=\nu\,C_0\,\veps_{AB}\veps_{cj}\,H_2^j,$$
\be
                                                            \label{calH}
{{\cal H}_{ABc}}^{(1)}= \nu\,\veps_{AB}\veps_{cj}\,e^{-3\xi_4}\,\Bigl( 2\,
{{H_2}^{j f}}_f
-{2\over 3}\,{{H_1}^j}_f\,{H_2}^f+{2\over 3}\,{{H_2}^j}_f\,{H_1}^f+
\ee

$$+{4\over 3}\,{{H_2}^f}_f\,{H_1}^j+ {2\over 3}\,{{H_1}^f}_f\,{H_2}^j+
{2\over 3}\,{H_2}^j\,({H_1}^f\,{H_1}_f) -
{2\over 3}\,{H_1}^j\,({H_1}^f\,{H_2}_f) \Bigr).
$$

${\cal H}_{\tilde A\tilde B c}$ can be written from (\ref{calH}) with the
 following interchange:
$$ \veps_{AB} \ra \veps_{\tilde A \tilde B},
\ \ \ \nu \ra \tilde \nu,
\ \ \ H_2\,\, \leftrightarrow
 \,\,H_1.$$

Using (\ref{calH}) and (\ref{A1}), (\ref{A2})
axionic EM  (eq. \ref{dHabc})
results in  2 equations:

 \be
\label{result2}
 \,\veps_{AB}\,\veps_{cd}\,
{\left( e^{-H_2}\,C_0 -2\,\alpha'\,{{H_2}_f}^f\,e^{H_1} \right)^{j}}_{j}=0,
\ee

 \be
\label{result1}
\, \veps_{\tilde A \tilde B}\,\veps_{cd}\,
{\left( e^{-H_1}\,C_0 -2\,\alpha'\,{{H_1}_f}^f\,e^{H_2} \right)^{j}}_{j}=0.
\ee

Let's go to the dilaton equation of motion. It can be presented in the form:

$${D_{\hat F}}^2 \phi +{1\over12}\phi  M^2_{\hat A\hat B\hat C}
-\alpha'\left[ -{2\over 3}(R_{\hat A\hat B})^2
 +{1\over3}(R_{\hat A\hat B\hat C\hat D})^2 -\right.$$
$$-{1\over 6}R^{\hat A\hat B} M^2_{\hat A\hat B}
  -{1\over18} M^{\hat A\hat B\hat C}{D_{\hat F}}^2 M_{\hat A\hat B\hat C}+
{1\over3} M^{\hat A \hat B \hat C;\hat D} M^2_{\hat A\hat B\hat C\hat D}-$$
\be
                                                               \label{dil}
\left. -{1\over24}  M^2_{\hat A\hat B\hat C\hat D}
M^{2  \hat A\hat C\hat B\hat D}
 -{1\over12} {D_{\hat F}}^2  M^2 +
{1\over6} ( M^2_{\hat A\hat B})^{;\hat A\hat B} \right] =0.
\ee

Calculation of  all the terms in eq.(\ref{dil}) with the help of relations
obtained before  gives:

\be
e^{H_1}\,{\left( e^{-H_1}\,C_0 -2\,\alpha'\,{{H_1}_f}^f\,e^{H_2}
 \right)^{j}}_{j} + e^{H_2}\,{\left( e^{-H_2}\,C_0 -2\,
\alpha'\,{{H_2}_f}^f\,e^{H_1}
 \right)^{j}}_{j}=0,
\ee
which is satisfied simultaneously with eq.(\ref{result2}), (\ref{result1}).

A  study of a rather complicated graviton equation of
motion ($\phi R_{\hat A\hat B} + \ldots =0 $) does not produce  additional
constraints for vacuum configuration.

\section{The choice of the solution}

One  must solve  eq.(\ref{result2}), (\ref{result1}), matching the
singular sources, the $1^{st}$   brane being the source
for eq.(\ref{result1}), the $2^{d}$ - for  eq.(\ref{result2}).
We want to discuss the solutions of the two types:
\begin{enumerate}
\item After introducing complex variable $z=y^8+i\,y^9$,
one can easily see, that
 equations (\ref{result2}), (\ref{result1}) are
automatically satisfied for arbitrary analytical functions:

$ H_1=H_1(z)$, $ H_2=H_2(z) $. Moreover, all $\alpha'$ corrections also
vanish in this case and we  have:
$ \phi = e^{-{1 \over 3}\,(H_1+H_2)},$  putting here and bellow $C_0=1$,
$${\cal H}_{67z}=\phi\,M_{67z}=i\,\tilde \nu \partial_z\,(e^{-H_1}),
\ \ \ {\cal H}_{67 \bar z}=0,\ \ \
{\cal H}_{45z}=\phi\,M_{45z}=i\, \nu \partial_z\,(e^{-H_2}),
\ \ \ {\cal H}_{45 \bar z}=0, $$
where $M_{PQz}={1 \over 2}\,( M_{PQ8}-i\,M_{PQ9}),$ etc.
We ought to choose:
\be
\label{sol1}
e^{-H_1}=C_1-q_1\,\ln {z \over {\rho}},\ \ \ 
e^{-H_2}=C_2-q_2\,\ln {z \over {\rho}}, \,\,\,q_1>0,\,\,\, 
q_2>0,\,\,\,C_1>0,\,\,\,C_2>0
\ee
to obtain non-zero charges:
\be
\label{ch1}
Q_1={1 \over 2\,\pi}\,\int \limits_{{|z|}^2={\rho}^2} {\cal H}_{67z}dz=
\tilde \nu \, q_1,\ \ \ \ \
Q_2={1 \over 2\,\pi}\,\int \limits_{{|z|}^2={\rho}^2} {\cal H}_{45z}dz=
 \nu \, q_2.
\ee
Note that integration of ${\cal H}^{(3)}$ over 3-sphere
$x_6^2+x_7^2+{|z|}^2={\rho}^2$
reduces to integration over circle ${|z|}^2={\rho}^2$, since
${\cal H}_{\tilde P \tilde Q r}$ is in fact 
$\sim \delta (x_6)\,\delta (x_7)$.
( Remember that $1^{\underline{st}}$ brane is situated in the origin 
of 6,7-coordinates. )
Analogously ${\cal H}_{PQr}\,\sim \delta (x_4)\,\delta (x_5)$.

The explicit formulas for dilaton and axion of the constructed solution are:
$$\phi=\Bigl (C_1-q_1\,\ln{z \over {\rho}}\Bigr )^{{1 \over 3} }\,
\Bigl (C_2-q_2\,\ln{z \over {\rho}}\Bigr )^{{1 \over 3} },$$
\be
\label{CC}
C_{012345}=\tilde \nu \Bigl (C_1-q_1\,\ln{z \over {\rho}}\Bigr )^{-1}, \ \ \
C_{012367}=\nu \Bigl (C_2-q_2\,\ln{z \over {\rho}}\Bigr )^{-1}.
\ee
 ( We must demand $ |z|<{\rho}$ to avoid singularities at $z \ne 0$ ).

The only non-zero component of Richi-tensor is:
\be
\label{Rzz}
R_{zz}={1 \over 9\,z^2}\,\Bigl [{ \Bigl (q_1\,e^{H_1} +
q_2\,e^{H_2}\Bigr )}^2+
 {3\over 2}\,q_1\,e^{H_1}\,\Bigl (5q_1\,e^{H_1}-2\Bigr )  +
{3\over 2}\,q_2\,e^{H_2}\,\Bigl (5q_2\,e^{H_2}-2\Bigr ) \Bigr ].
\ee

Curvature  scalar R  and lagrangian ( eq.\ref{1} ) are  zero on 
this solution ( it is true for any $H_1(z)$ and $H_2(z)$ ).

It is worth emphasizing, that if we  had neglected $\alpha'$ corrections,
then the general solution   would have been:

\be
\label{er1}
e^{-H_1}=C_1-q_1\,\ln {z \over {\rho}} -
q_1{}^{\prime}\,\ln {\bar z \over {\rho}}, \,\,\,q_1 \ge 0,
q_1{}^{\prime} \ge 0;
\ee
 \be
\label{er2}
e^{-H_2}=C_2-q_2\,\ln {z \over {\rho}} -
q_2{}^{\prime}\,\ln {\bar z \over {\rho}},
\,\,\,q_2 \ge 0, q_2{}^{\prime} \ge 0.
 \ee

If  one of the $q_1$, $q_1{}^{\prime}$ and one of the
 $q_2$, $q_2{}^{\prime}$ is set to zero, 
the choice (\ref{er1}), (\ref{er2}) goes through
eq.(\ref{result2}), (\ref{result1}) without any modifications,
but this is not the case if  $q_1 \ne 0$, $q_1{}^{\prime} \ne 0$ or
 $q_2 \ne 0$, $q_2{}^{\prime} \ne 0$.

\item  Let us analyze the solution of the type $H_1(r)$,
$H_2(r)$, possessing SO(2) symmetry in z-plane, here $r={|z|}^2$.
We integrate  eq.(\ref{result2}), (\ref{result1}) twicely and write:
\be
\label{r1}
e^{-H_1} -2\,\alpha'\,{{H_1}_f}^f\,e^{H_2}=C_1- q_1\,\ln{r \over {\rho}},
\ee
\be
\label{r2}
e^{-H_2} -2\,\alpha'\,{{H_2}_f}^f\,e^{H_1}=C_2 - q_2\,\ln{r \over {\rho}}.
\ee
One can see, that in the vicinity of $r=0$ the solution of this system
 behaves as :
\be
\label{Z1Z2}
 e^{-H_1} \sim -\sqrt{2\,\alpha'}\,
{\Bigl (r \ln {r \over {\rho}} \Bigr )}^{-1} , \ \ \
e^{-H_2} \sim -\sqrt{2\,\alpha'} \,
{\Bigl (r \ln {r \over {\rho}} \Bigr )}^{-1}.
\ee

In the limit $r \ra 0$ it is more
singular than $\ln {r \over {\rho}} $, which would have described 
the behaviour of the solution, if we had neglected terms with $\alpha'$ 
in (\ref{r1}), (\ref{r2}).
For (standard) dilaton and curvature scalar in  $r \ra 0$ limit we obtain:
\be
\label{varphi}
 e^{2\,\varphi}={{4^2\,\alpha'} \over {9^3\,r^2}}\,
{\Bigl (\ln {r \over {\rho}} \Bigr )}^4,\ \ \ \  R=-{1\over r^2}.
\ee

Far from the origin one can solve eq.(\ref{r1}), (\ref{r2})
by expanding in $\alpha'$ 
( assuming $\alpha'\,\ln\Bigl ({r \over {\rho}}\Bigr )\ll 1$)
 and using  $q_1 \sim\,\alpha'$,
 $q_2 \sim\,\alpha'$ ( see \cite{Call1}, \cite{Town} ), so that the
expressions for the defining functions are:
\be
\label{sol21}
e^{-H_1}=C_1-q_1\,\ln {r \over {\rho}} - 2\alpha'\,
{q_1^2 \over {C_1^2\,C_2\,r^2}} + O\Bigl ({\alpha'}^4\Bigr ),\ \  C_1>0;
\ee
\be
\label{sol2}
e^{-H_2}=C_2-q_2\,\ln {r \over {\rho}} - 2\alpha'\,
{q_2^2 \over {C_2^2\,C_1\,r^2}} + O\Bigl ({\alpha'}^4\Bigr ),\ \  C_2>0.
\ee

The SO(2) symmetric  solution has charges:
\be
\label{ch11}
Q_1= {1 \over 2\,\pi}\int \limits_{{ |z| }^2=
{\rho}^2} (-\tilde \nu \partial_r\,e^{-H_1}\Bigr )\,dz=
\tilde \nu \, q_1,\ \
Q_2= {1 \over 2\,\pi}
\int \limits_{{|z|}^2={\rho}^2} (-\nu \partial_r\,e^{-H_2}\Bigr ) dz=
 \nu \, q_2.
\ee
Here only ${\cal H}_{67m}^{(0)}$ ( ${\cal H}_{45m}^{(0)}$ )
 contributes to $Q_1$ ( $Q_2$ ), since ${\cal H}_{67m}^{(1)}$,
${\cal H}_{45m}^{(1)}$ vanish at infinity more rapidly than $ {1 \over r}$.

We give also the formula for dilaton up to  $O \Bigl ({\alpha'}^4 \Bigr )$:
\be
\label{phi3}
e^{2\,\varphi}=C_1\,C_2 - (q_1\,C_2 + q_2\,C_1)\,
\ln {r \over {\rho}} + q_1\,q_2 \,
{\Bigl ( \ln {r \over {\rho}} \Bigr)}^2 + {\alpha' \over 3\,r^2}\,
{\Bigl ( {q_1\over C_1} +  {q_2\over C_2} \Bigr )}^2.
\ee

\end{enumerate}
\section{Conclusion}
Let's summarize our results. We've constructed the solution of
anomaly-free N=1, D=10 Dual Supergravity, which corresponds to the presence
of the two intersecting 5-branes. We've found that accounting for
higher-order-derivative ( up to the $4^{(th)}$ ) terms in the Lagrangian
bring forth modifications of the solution. An important thing is that 
the two functions $H_1$ and $H_2$, determing all the fields, 
are mixed in eq.(\ref{result2}), (\ref{result1}).
Let us recall that the solutions of Lagrangians with up to the $2^d$
order-derivative-terms, widely discussed in literature, are characterized
by the independent harmonic functions (\cite{Aref}, \cite{Pope},
\cite{Arg}). $e^{-H_1}$, $e^{-H_2}$ would have
played the role of these functions in the solution without
 $\alpha'$ corrections,

In the SO(2) symmetric case modifications display themselves not only in
 $\alpha'$ corrections to the solution far from the origin, but also
in changing the type of singular behaviour in the vicinity of $r=0$.

 Remarkably,  analytic solution is  exact: field configuration doesn't
acquire $\alpha'$ modifications and $H_1$, $H_2$ remain independent.

\vspace{1cm}

The author would like to thank A.S.Gorsky and K.N.Zyablyuk for
useful discussions.
\newpage
\section{Appendix}

Here are the formulas for non-zero components of 10D-spin-connection:

\be
\label{A1}
{\omega_{\alpha\beta}}^ c = 
-\,e^{-\xi_4}\,\eta_{\alpha\beta}\,{\xi_1}^c,\ \ \
{\omega_{AB}}^c=-e^{-\xi_4}\,\eta_{AB}\,{\xi_2}^c,\ \ \
{\omega_{\tilde A \tilde B}}^c=
-e^{-\xi_4}\,\eta_{\tilde A\tilde B}\,{\xi_3}^c,
\ee

\be
\label{A2}
\omega_{abc}=-2e^{-\xi_4}\,\eta_{a[b}\,{\xi_2}_{c]},
\ee
where
$\eta_{\hat A \hat B}=(1,-1,-1,-1,-1,-1,-1,-1,-1,-1).$

The folowing  curvature tensor componets  survive
($R =d\omega + \omega \wedge \omega $):

$${R_{\alpha\beta}}^{\gamma \delta}=
2\,e^{-2\,\xi_4}\,\delta^{\,\gamma}_{[\alpha}
\delta^{\,\delta}_{\beta]}\,{\xi_1}^f{\xi_1}_f \ \ \
{R_{AB}}^{CD}=
2\,e^{-2\,\xi_4}\,\delta^{\,C}_{[A}
\delta^{\,D}_{B]}\,{\xi_2}^f{\xi_2}_f $$

$${R_{\alpha A}}^{\beta B}=
e^{-2\,\xi_4}\,\delta^{\,\beta}_{\alpha}
\delta^{\,B}_{A}\,{\xi_1}^f{\xi_2}_f \ \ \
{R_{\tilde A \tilde B}}^{\tilde C \tilde D}=
2\,e^{-2\,\xi_4}\,\delta^{\,\tilde C}_{[\tilde A}
\delta^{\,\tilde D}_{\tilde B]}\,{\xi_3}^f{\xi_3}_f$$
\be
\label{A3}
{R_{\alpha \tilde A}}^{\beta \tilde B}=
e^{-2\,\xi_4}\,\delta^{\,\beta}_{\alpha}
\delta^{\,\tilde B}_{\tilde A}\,{\xi_1}^f{\xi_3}_f \ \ \
{R_{A \tilde B}}^{C \tilde D}=
e^{-2\,\xi_4}\,\delta^{\,C}_{A}
\delta^{\,\tilde D}_{\tilde B}\,{\xi_2}^f{\xi_3}_f
\ee

$$ {R_{\alpha b}}^{\gamma c}=e^{-2\,\xi_4} \delta_\alpha^\gamma\,
({{\xi_1}_b}^c -\,{\xi_1}_b{\xi_4}^c-\,{\xi_4}_b{\xi_1}^c+
{\xi_1}_b{\xi_1}^c + \delta_b^c\, {\xi_1}^f{\xi_4}_f) $$

$$ {R_{A b}}^{C d}=e^{-2\,\xi_4} \delta_A^C\,
({{\xi_2}_b}^d -\,{\xi_2}_b{\xi_4}^d-\,{\xi_4}_b{\xi_2}^d+
{\xi_2}_b{\xi_2}^d + \delta_b^d\, {\xi_2}^f{\xi_4}_f) $$

$$ {R_{\tilde A b}}^{\tilde C d}=e^{-2\,\xi_4} 
{\delta_{\tilde A}}^{\tilde C}\,
({{\xi_3}_b}^d -\,{\xi_3}_b{\xi_4}^d-\,{\xi_4}_b{\xi_3}^d+
{\xi_3}_b{\xi_3}^d
 + \delta_b^d\, {\xi_3}^f{\xi_4}_f) $$

$${R_{ab}}^{cd}=e^{-2\xi_4}\,( 4\,
\delta_{[a}{}^{[c} \,{\xi_4}_{b]}{}^{d]} -
4\,\delta_{[a}{}^{[c} \,{\xi_4}_{b]}{\xi_4}^{d]} +
2\,\delta_{[a}^{\,c}\delta_{b]}^d\,
{\xi_4}^f{\xi_4}_f)$$


\begin{thebibliography}{99}
\bibitem{Dabh} \ A.Dabholkar, G. Gibbons, J.A.Harvey and F.R.Ruiz, \\
 \NP B 340 ( 1990 ) 33.
\bibitem{Strom} \ A.Strominger, \NP B 343 (1990)167.
\bibitem{Duff1} \ M.J.Duff and J.X.Lu,
\NP B 354 (1991)141.
\bibitem{Call1} \ C.G.Callan, J.A.Harvey, A.Strominger,
\NP B 359(1991)611.
\bibitem{Call2} \ C.G.Callan, J.A.Harvey, A.Strominger,
 \NP B 367 ( 1991 )60.
\bibitem{Duff2}  \ M.J.Duff and J.X.Lu,
 \NP B 416 ( 1994 )301.
\bibitem{Duff3} \ M.J.Duff and R.Minasian,
 \NP B 436 (1995)507.
\bibitem{Town} \ P.K.Townsend, hep-th 9507048.
\bibitem{Schw} \ John H.Schwarz, hep-th/9607201.
\bibitem{Stelle} \ K.S.Stelle, hep-th/9701088.
\bibitem{Dougl} \  M.Berkooz, M.Douglas, R.Leigh, hep-th/9606139.
\bibitem{Aref} \ I.Ya. Aref'eva, M.G.Ivanov, O.A.Rytchkov,
 hep-th/9702077.
\bibitem{Pope} \ C.N. Pope, H. Lu,
hep-th/9702086.
\bibitem{Gaunt} \ J.P.Gauntlett, G.W.Gibbons, G.Papadopoulos and\\
P.K.Townsend, hep-th/9702202.
\bibitem{Arg} \ R.Argurio, F.Englert, L.Houart,
 hep-th 9701042.
\bibitem{Tseyt} \ A.A.Tseytlin,  hep-th 9702163.
\bibitem{Roo} \ M.de Roo , hep-th 9703124.
\bibitem{D}\ M.J.Duff, \ \ Class. Quant. Grav. {\bf 5} (1988) 189.
\bibitem{DL}\ M.J.Duff, J.X.Lu  \ \ Phys. Rev. Lett.{\bf 66} (1988) 1402,\\
Class. Quant. Grav/ {\bf 9} (1991) 1.
\bibitem{STZ1} \ N.A.Saulina,  M.V.Terentiev, K.N.Zyablyuk, \\
~~\PL~~{\bf B 366} (1996) 134.
\bibitem{STZ2} \ N.A.Saulina,  M.V.Terentiev, K.N.Zyablyuk, \\
 Intern. J. Mod. Phys.{\bf A}. ( accepted to publication in
Sep. 1997 )
\bibitem{TZ} \ M.V.Terentiev, K.N.Zyablyuk, unpublished
\bibitem{CH} \ A.Chamseddine, ~~Phys. Rev. {D 24} (1981) 3065.
\bibitem{CM}\ E.Chapline, N.Manton,~~\PL~~{\bf B 120} (1983) 105.
 \bibitem{GS}\ M.Green, J.Schwarz,~~\PL~~{\bf B 149} (1984) 117.
\bibitem{BBLPT} \  L.Bonora, M.Bregola, K.Lechner, P.Pasti, M.Tonin,\\
 ~~\NP ~~{\bf B 296} (1988) 877.
\bibitem{AFRR}\ R.D'Auria, P.Fre,
M.Raciti, F.Riva, ~~Intern. J. Mod. Phys.  {\bf A 3}\ (1988)\ 953.
\bibitem{P} \ I.Pesando, ~~Phys. Lett. {\bf B 272} (1991) 45, \\
Class. Quantum Grav. {\bf 9} (1992) 823.

\end{thebibliography}
\end{document}